\documentclass[aps,prl,twocolumn,groupedaddress]{revtex4}
\usepackage{graphicx}
\begin{document}

\title{Understanding the problem of glass transition on the basis of elastic interactions in a liquid}
\author{Kostya Trachenko$^{1}$}
\address{$^1$ Department of Earth Sciences, University of Cambridge, UK}

\begin{abstract}
We review the recently proposed elastic approach to glass transition. This approach is based on a simple and a physically transparent idea of elastic interactions between local relaxation events in a liquid. Central to this picture is the range of this interaction. Its increase on lowering the temperature explains several important open questions in the area of glass transition, including universal relaxation laws and dynamic crossovers. In particular, we show how the proposed theory explains (1) the physical origin of cooperativity of relaxation; (2) the origin of the crossover from exponential to non-exponential relaxation at $\tau=$1 ps, where $\tau$ is liquid relaxation time; (3) the origin of the Vogel-Fulcher-Tammann law; (4) the origin of stretched-exponential relaxation; (5) the absence of divergence of $\tau$ at the VFT temperature $T_0$ and the crossover to a more Arrhenius relaxation at $\tau\approx 10^{-6}$ sec; (6) the origin of liquid ``fragility''; and (7) the relationship between non-exponentiality of relaxation and relaxation time.
\end{abstract}


\maketitle

The problem of glass transition has been widely discussed, and is considered to be one of the main enduring controversies in physics \cite{langer,dyre,angell,ngai,angell1,phillips}. As widely perceived, a glass transition theory should provide a consistent explanation of several universal relaxation properties of liquids which set in on lowering the temperature, including the physical origin of cooperativity of relaxation, slow non-exponential dynamics, the Vogel-Fulcher-Tammann law, dynamic crossovers and other phenomena \cite{langer,dyre,angell,ngai,angell1,phillips,sti,cross,casa,cross2}.

Several current theories of glass transition proposed new, often exciting, physical mechanisms. These mechanisms are governed by
different parameters that control glass transition: free volume, entropy, energy landscape, mode coupling and others \cite{dyre}. Part of the challenge for a consistent theory is that there is no agreement as to what is the relevant parameter that controls
glass transition, or whether it exists at all \cite{dyre}.

We have recently asked whether glass transition can be understood solely on the basis of elasticity of a liquid \cite{e1,e11,e3,e2}. Such a question is important from at least two perspectives. First, if answered positively, we would lose the need for some ad-hoc assumptions and postulates \cite{dyre}. Because these are not based on transparent physics, they often obscure the view of glass transition \cite{dyre}. Second, it is important to discuss elasticity because a glass is different from a liquid by its ability to support static shear stress. Hence stress relaxation, or elasticity, should feature as an important, preferably as a starting, point of a discussion of glass transition.

With this reasoning in mind, we have developed an elastic approach to glass transition. This approach is based on a simple and physically transparent idea: {\it local relaxation events in a liquid interact via the elastic fields they induce}. The important parameter here is the range of this interaction, which we called the ``liquid elasticity length'' $d_{\rm el}$. We proposed that the most important open questions in the area of glass transition \cite{langer,dyre,angell,ngai,angell1,phillips} can be understood by discussing how $d_{\rm el}$ changes with temperature \cite{e1,e11,e3,e2}. In this paper, we briefly review this work. We discuss (1) the origin cooperativity of relaxation; (2) crossover from exponential to non-exponential relaxation at $\tau=$1 ps, where $\tau$ is liquid relaxation time; (3) the origin of the VFT law; (4) the origin of stretched-exponential relaxation; (5) the absence of divergence at the VFT temperature $T_0$ and the crossover to a more Arrhenius relaxation at $\tau\approx 10^{-6}$ sec; (6) the physical origin of liquid ``fragility''; and (7) the relationship between non-exponentiality of relaxation and relaxation time.

We begin the discussion with local relaxation events (LREs) which give flow in a liquid. A LRE involves the jump of an atom from
its cage and subsequent relaxation of the local structure (see Figure 1). The activation barrier for a LRE at the constant volume of the cage is very large due to strong short-range interatomic repulsions. Hence, during a LRE, some of the cage atoms should move outwards, increasing volume of the cage \cite{dyre}. In doing so, the work is performed to deform the surrounding liquid, which sets the activation barrier for a LRE. The deformation of the liquid outside the cage does not involve, to the first order, density change, i. e. can be considered as a shear deformation. As discussed by Dyre \cite{dyre}, this is because the radial displacement $u_r$ from an expanding sphere decays as $u_r\propto 1/r^2$ \cite{elast}, giving div$(u_r)=0$. Therefore, in discussing how LREs interact elastically, we consider shear LREs.

\begin{figure}
\begin{center}
{\scalebox{0.4}{\includegraphics{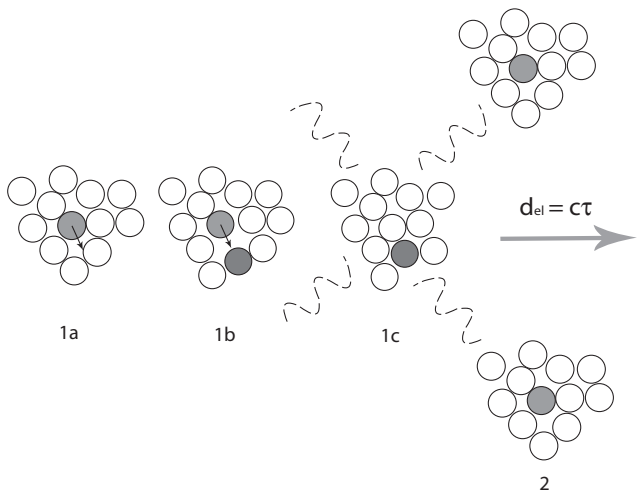}}}
\end{center}
\caption{Illustration of the elastic interaction between local relaxation events. Jump of the atom (1a) involves deformation of its cage (1b), resulting in shear elastic waves propagating from this event. These waves deform the cages of remote local regions (2) and affect their relaxation. The elastic interaction between LREs operates in the range $d_{\rm el}=c\tau$.}
\end{figure}

Lets consider how a given local relaxing region (local region 2 in Figure 1) is affected by stresses due to remote shear LREs originating from, say, region 1 in the same Figure. Lets assume that region 2 is located in the centre of a sphere. Relaxation of the central event involves deformation of the ``cage'' around the jumping atom, and therefore depends on the stresses that propagate from the remote LREs to the centre. A remote shear LRE creates elastic shear waves, which include waves of high frequency. This is because the deformation, associated with a LRE, creates a wave with a length comparable to interatomic separations, and hence with a frequency on the order of the Debye frequency. At high frequency $\omega>1/\tau$, a liquid supports propagating shear waves \cite{frenkel}, which propagate stress and its variations from remote LREs to the central point. If $\tau$ is macroscopically defined as the time of decay of shear stress in a liquid \cite{frenkel,elast}, $d_{\rm el}=c\tau$ gives the length of this decay, where $c$ is the speed of sound. Here, $d_{\rm el}$ gives an estimation of the maximal range over which shear stress decays in a liquid. At the microscopic level, the relevance of $d_{\rm el}=c\tau$ is as follows. A high-frequency shear wave originating from a LRE propagates stress until a remote LRE takes place at the front of the wave, at which point the wave front is absorbed by the remote LRE. Suppose this happens at distance $d_{\rm el}$ from the original LRE. $d_{\rm el}$ can be calculated from the condition of equality of the wave travel time, $d_{\rm el}/c$, and the time at which the remote LRE takes place at point $d_{\rm el}$. The latter time is given by $\tau$, because microscopically, $\tau$ is defined as the average time between two consecutive LREs at one point in space \cite{frenkel}, and we obtain $d_{\rm el}=c\tau$ as before.

Therefore, $d_{\rm el}$ defines the maximal distance over which a given (central) LRE is affected by elastic shear stresses due to other LREs in a liquid. For this reason, $d_{\rm el}$ can be called the {\it liquid elasticity length}. Note that relaxation of the central event is affected by all those stresses that have enough time to propagate to the centre. Because it takes time $\tau$ for the central event to relax, its relaxation is affected by the stresses from all LREs located distance $c\tau$ away. After time $\tau$, the central event relaxes, and the process repeats. Therefore, the definition $d_{\rm el}=c\tau$ is self-consistent.

Because $c$ is on the order of $a/\tau_0$, where $a$ is the interatomic separation of about 1 \AA\ and $\tau_0$ the oscillation
period, or inverse of Debye frequency ($\tau_0\approx 0.1$ ps),

\begin{equation}
d_{\rm el}=c\tau=a\frac{\tau}{\tau_0}
\end{equation}

On lowering the temperature, $\tau$ increases as $\tau=\tau_0\exp(V/kT)$, where $V$ is the activation barrier of a LRE \cite{frenkel} (here, $V$ can be temperature-dependent). According to Eq. (1), this increases $d_{\rm el}$ and the number of LREs that elastically interact with a given event. We have proposed \cite{e1} that this is the key to understanding glass transition. As discussed below, this simple and physically transparent idea explains many important open questions in the area of glass transition.

{\bf 1. The origin of cooperativity of relaxation}

One of the central problems in the area of glass transition is the physical origin of ``cooperativity''. The notion of cooperativity of molecular motion, which sets in a liquid as temperature is lowered, was introduced and intensely discussed in several popular theories of the glass transition. These theories are based on the assumption that ``cooperatively rearranging regions'', ``domains'' or ``clusters'' exist in a liquid, in which atoms move in some concerted way that distinguishes these regions from their surroundings \cite{angell,dyre,langer,ngai,angell1,yama,adam}. The physical origin of cooperativity is not understood, nor is the nature of concerted motion.

Eq. (1) immediately explains the origin of cooperativity of relaxation. When, at high temperature, $\tau\approx\tau_0$, $d_{\rm el}\approx a$ (see Eq. (1)). Hence, $d_{\rm el}<d_m$, where $d_m$ is the distance between neighbouring LREs of about 10 \AA\ ($d_m$ is the distance between the centres of neighbouring molecular cages). This means that LREs do not elastically interact. As $\tau$ increases on lowering the temperature, $d_{\rm el}\ge d_m$ becomes true. At this point, LREs are no longer independent, because relaxation of a LRE is affected by elastic stresses from other events. This discussion, therefore, clarifies the physical origin of cooperativity. Here, we do not need to assume or postulate cooperativity of relaxation as in the previous work \cite{angell,dyre,langer,ngai,angell1,yama,adam}. In this picture, relaxation is ``cooperative'' in the general sense that LREs are not independent, but the origin of this cooperativity is the usual elastic interaction.

{\bf 2. The origin of the crossover from exponential to non-exponential relaxation at $\tau=$1 ps}

As immediately follows from point (1), the crossover from exponential to non-exponential relaxation takes place when $d_{\rm el}=d_m$. According to Eq. (1), $\tau$ at the crossover, $\tau_c$, is a universal value: $\tau_c=\tau_0 d_m/a$. This gives $\tau_c$ of about 1 ps, consistent with the numerous experiments \cite{cross,casa}.

{\bf 3. The origin of the Vogel-Fulcher-Tammann law}

At high temperature, relaxation time $\tau$ of a liquid follows Arrhenius dependence. On lowering the temperature, $\tau$ almost universally deviates from Arrhenius dependence, and follows the Vogel-Fulcher-Tammann (VFT) law, $\tau=\tau_0\exp\left(\frac{A}{T-T_0}\right)$, where $A$ and $T_0$ are constants. The origin of the VFT law is the main open question in the field of the glass transition \cite{langer,dyre}.

In order to derive the VFT law, we recall that $V$ is given by the work of the elastic force needed to deform the liquid around a LRE, i.e. by the liquid elastic energy \cite{dyre,nemilov,dyre1}. Because this deformation does not result in the compression of the surrounding liquid (for the deformation field $\bf u_r$, div$(\bf u_r)=0$), $V$ is given by the background shear energy of the liquid. This was confirmed by the experiments showing that $V$ increases with the liquid shear energy \cite{dyre1}.

As $d_{\rm el}$ increases on lowering the temperature, a given local region elastically interacts with an increasing number of other LREs in a liquid. This increases the background shear elastic energy of the liquid around the currently relaxing region and, therefore, the activation barrier of its LRE, $V$. The contributions of stresses due to remote LREs can be integrated by noting that elastic stresses decay as $\propto 1/r^3$ \cite{elast} and that the number of local regions in the sphere increases as $\propto r^2$. This gives the result that $V$ increases as a logarithm of $d_{\rm el}$:

\begin{equation}
V=V_0+C\ln(2d_{\rm el}/d_0)
\end{equation}

\noindent where $V_0$ is the high-temperature (non-cooperative) activation barrier, $d_0$ is the size of local relaxing region of about 10 \AA\ and constant $C$ is given by other microscopic parameters \cite{e1}.

Using $\tau=\tau_0\exp(V/kT)$ in Eq. (1), we obtain

\begin{equation}
d_{\rm el}=a\exp\left(\frac{V}{kT}\right)
\end{equation}

Eqs. (2) and (3) define $V$ in a self-consistent way. Eliminating $d_{\rm el}$ from the two equations, we find:

\begin{equation}
V=\frac{AT}{T-T_0}
\end{equation}

\noindent where $A=V_0+C\ln(2a/d_0)$ and $kT_0=C$.

From Eq. (4), the VFT law follows. In this picture, the super-Arrhenius behaviour is related to the increase of $d_{\rm el}$ (see Eq. (2)). The transition from the VFT law to the Arrhenius form of $\tau$ takes place in the limit of small $d_{\rm el}$ at high temperature. In this case, $V=V_0$ and $\tau=\tau_0\exp(V_0/kT)$.

{\bf 4. The origin of stretched-exponential relaxation}

At high temperature, the response of a liquid to an external perturbation (e.g., pressure or field) is exponential, $q(t)\propto\exp(-t/\tau)$, where $q(t)$ is a relaxing quantity. On lowering the temperature, the response deviates from exponential, and becomes stretched-exponential: $q(t)\propto\exp(-(t/\tau)^\beta)$, where $\beta$ is the non-exponentiality parameter, $0<\beta\le 1$ \cite{angell,dyre,ngai,angell1,phillips}. Stretched-exponential relaxation (SER) describes a very slow dynamics: in the wide range, it decays as a logarithm of time. Similar to the VFT law, the origin of SER is one of the most important open questions in the area of glass transition \cite{angell,dyre,ngai,angell1,phillips}. It is also the oldest problem in the area \cite{phillips}.

In the proposed theory to glass transition, SER can be derived as follows. Lets consider relaxation of a liquid in response to external perturbation (e.g. pressure or field). We introduce time-dependent variable $n(t)$, the current number of LREs induced by an external perturbation in a liquid in the sphere of radius $d_{\rm el}$. We consider relaxation at constant temperature, which fixes $d_{\rm el}$. $n(t)$ starts from zero and increases to its final value $n_{\rm r}$, at which point relaxation to equilibrium is complete. Lets consider the current LRE to relax to be in the centre of the sphere of radius $d_{\rm el}$. As discussed above, all previous remote LREs within distance $d_{\rm el}$ from the centre participate in the elastic interaction with the central event, changing the background shear energy of the liquid around the central event and hence affecting its activation barrier $V$. We have called this process ``elastic feed-forward interaction mechanism'' \cite{e1,e2}. By accounting for the decay of elastic stresses in the sphere of radius $d_{\rm el}$ in a similar way as in Eq. (2), one finds the activation barrier for the current LRE, $V(n)=V_{\rm VFT}+V_1\frac{n}{n_r}$, where $V_{\rm VFT}$ is set by the elastic interactions in an equilibrium liquid and is given by Eq. (4), $V_1=C_1\ln(2d_{\rm el}/d_0)$, and $C_1$ is a constant \cite{e2}. The rate of LREs, ${\rm d}n/{\rm d}t$, is the product of the number of unrelaxed events, $(n_{\rm r}-n)$, and the event probability, $\rho=\exp(V(n)/kT)$. Introducing $q=n/n_r$ and reduced time $t/\tau_0$, we write:

\begin{equation}
\frac{{\rm d}q}{{\rm d}t}=(1-q)\exp\left(-\frac{V_{\rm VFT}+V_1q}{kT}\right)
\end{equation}

Eq. (5) has two parameters, $V_{\rm VFT}/kT$ and $\alpha=V_1/kT$. We solved Eq. (5) for a range of these parameters, and showed that the solution is described by SER, $q(t)=1-\exp{(-(t/\tau)^\beta})$ \cite{e2}. Note that whereas $\tau$ depends on both $V_{\rm VFT}/kT$ and $\alpha$, $\beta$ depends on $\alpha$ only ($\alpha=0$ gives exponential relaxation, $\beta=1$). Hence, Eq. (5) is, in fact, a one-parameter equation for $\beta$: $\frac{{\rm d}q}{{\rm d}t}=(1-q)\exp(-\alpha q)$. Consistent with the experiments \cite{angell,dyre,ngai,angell1}, we find that $\beta$ decreases from 1 at high temperature to smaller values as the temperature is lowered \cite{e11,e2}.

The transition from SER to exponential relaxation takes place in the limit of small $d_{\rm el}$ at high temperature. In this case, $V_1=0$, $V=V_0$ and $\frac{{\rm d}q}{{\rm d}t}\propto (1-q)$, from which exponential relaxation follows.

{\bf 5. The absence of divergence at the VFT temperature $T_0$ and the crossover to a more Arrhenius relaxation at $\tau\approx 10^{-6}$ sec}

An important open question follows from the form of the VFT law, namely what happens at $T_0$. Because $\tau$ formally diverges at $T_0$, several models have suggested that a phase transition from a liquid to a glass phase can exist \cite{langer,dyre}. Because the divergence is not observed in an experiment, it was proposed that the phase transition is avoided due to sluggish dynamics when $\tau$ exceeds experimental time scale. However, the nature of the phase transition and the second phase is not clear, which continues to fuel the current debate \cite{langer,dyre,angell1}. Interestingly, the VFT law changes to a more Arrhenius form at low temperature, pushing the divergence temperature down \cite{sti}. The origin of this crossover is not understood.

In the proposed theory of the glass transition, the ongoing controversy regarding the divergence and possible phase transition at $T_0$ is readily reconciled. The divergence at $T_0$ can not exist for the following reason. From Eqs. (3,4), we find

\begin{equation}
d_{\rm el}=a\exp\left(\frac{A}{T-T_0}\right)
\end{equation}
\noindent

When $T$ approaches $T_0$, $d_{\rm el}$ diverges, and quickly exceeds any finite size of the system $L$. When $d_{\rm el}\ge L$, all LREs in the system elastically interact, and there is no room for the increase of $V$ by way of increasing $d_{\rm el}$. The upper limit of integral (3) becomes $d_{\rm el}=L$, giving temperature-independent $V\propto \ln(L)$ (see Eq. (2)). Further decrease of temperature has a weaker effect on $V$, and can be due to, e.g., density increase, but not to the increase of $d_{\rm el}$ (the density-related contribution to $V$ does not depend on $d_{\rm el}$ or $L$). As a result, the behaviour of $\tau$ tends to Arrhenius, pushing the divergence to zero temperature.

$d_{\rm el}$ exceeds the experimental value of $L$ above the glass transition temperature $T_g$: if $\tau(T_g)=10^3$ sec, $d_{\rm el}(T_g)=10^3$ km, according to Eq. (1). Hence our theory predicts the crossover from the VFT law to a more Arrhenius behaviour at low temperature, as is seen in the experiments \cite{sti}. According to Eq. (1), $\tau$ at the crossover is $\tau=\tau_0 L/a$. If a typical value of $L$ is 1 mm, $\tau$ at the crossover is $10^{-6}$ sec, consistent with the experimental results \cite{cross2}.

We note here that $d_{\rm el}$ vastly exceeds the size of ``cooperatively rearranging regions'' (CRR), which is several nm at $T_g$ (for review of CRR, see, e.g., Ref. \cite{yama}). The physical picture of CRR is not clear \cite{dyre}. It is possible that the observed nm scale of CRR is set by the distance beyond which the elastic strains from LREs decay to the values undistinguishable from thermal fluctuations.

{\bf 6. The physical origin of liquid ``fragility''}

The ``fragility'' plot has been, perhaps, the most common graph shown in papers and conferences dedicated to glass transition \cite{angell,angell1}. Its essence is the dependence of $V$ on temperature: in fragile liquids, $V$ significantly increases on lowering the temperature, whereas in strong liquids, $V$ is constant, or nearly constant. The physical origin of liquid fragility is not understood \cite{angell,dyre,angell1}.

In the proposed theory, the origin of fragility is easily understood on the basis of $d_{\rm el}$. According to Eq. (2), as long as at high temperature $d_{\rm el}<L$, lowering the temperature increases $V$, resulting in a fragile behaviour. If, on the other hand, $d_{\rm el}\ge L$ at high temperature already, further decrease of temperature has a weaker effect on $V$, giving strong behaviour. Experimentally, for many systems the studied range of temperatures varies from about $2T_g$ and $T_g$ \cite{casa}, hence we consider the increase of $d_{\rm el}$ from high temperature $T_h=2T_g$ to $T_g$. Take, for example, two systems on the stronger side of fragility plots, BeF$_2$ and SiO$_2$. From the experimental values of $V_h/kT_g$ ($V_h$ is the activation barrier at the highest measured temperature), we find $V_h/kT_h=24$ and 19.6 for BeF$_2$ and SiO$_2$, respectively \cite{novikov}. According to Eq. (3), this gives $d_{\rm el}=2.6$ m and 33 mm at $T_h$ for the two systems. Because a typical experimental value of $L$ is on the order of 1 mm, our theory correctly predicts that these systems should be on the strong end of fragility plots. For two fragile systems, toluene and propylene carbonate, $V_h/kT_h=3.34$ and 5.75, giving $d_{\rm el}=28$ and 314 \AA\ at $T_h$, respectively. This is much smaller than $L$, hence our theory predicts that these systems should be fragile, as is seen experimentally.

An interesting prediction from this picture is that strong systems will show increased fragility if the measurements are extended at high temperature so that $d_{\rm el}<L$ (note that strong systems have been measured at relatively low temperature only \cite{angell,angell1}).

{\bf 7. The relationship between non-exponentiality of relaxation $\beta$ and relaxation time $\tau$}

The non-exponentiality parameter $\beta$ and relaxation time $\tau$ are two fundamental parameters that describe a liquid in the glass transformation range. $\beta$ and $\tau$ have been discussed in a number of popular theoretical approaches \cite{e11}. A notable feature of these approaches is that $\beta$ and $\tau$ are often treated separately, and on the basis of unrelated physical mechanisms \cite{e11}. In view of this, it has remained unclear what the relationship between $\beta$ and $\tau$ is, or if one exists at all.

It is natural to ask whether both $\beta$ and $\tau$ are affected by the same slowing-down mechanism. In this case, a well-defined relationship between these parameters should exist. By analyzing the experimental data of 15 different glass-forming systems, we have recently shown that $\beta$ and $\tau$ are uniquely related in the glass transformation range \cite{e11}. First, at high temperature, $\log(\tau)$ is approximately proportional to $1/\beta$. Second, a crossover to another higher slope takes place at lower temperature \cite{e11}.

The key to the common relationship between $\beta$ and $\tau$ is that the two quantities are governed by the same parameter, $d_{\rm el}$ \cite{e11}. The increase of $d_{\rm el}$ on lowering the temperature increases $\tau$ (see Eq. (2)), and at the same time it decreases $\beta$ (see Eq. (5)). By solving Eq. (5) and joining the result with Eq. (1), we have shown that $\beta$ and $\tau$ are related in a way that is observed in the experiment \cite{e11}. The origin of crossovers in the slope of $\log(\tau)$ vs $1/\beta$ can be understood using the same reasoning as in point (5) above: the crossover takes place when $d_{\rm el}=L$.

Before concluding, we note that the properties discussed above are all related to the liquid behaviour above the glass transition temperature $T_g$. $T_g$ is defined from the condition of $\tau$ exceeding the experimental time scale of $100-1000$ sec. Hence, the transition of a liquid to a glass is not a mystery in itself \cite{dyre}: it universally takes place when $\tau$ exceeds the time of observation and a liquid falls out of equilibrium. However, a notable effect at $T_g$ is the change of the constant-pressure heat capacity, $\Delta c_p$. This has given rise to approaches that view glass transition as a second-order phase transition with a certain change of the system equilibrium property (e. g., entropy \cite{dyre,angell1,adam}). The entropy theory has been convincingly criticized for a number of reasons, including the problem of identifying the ``ideal'' glass phase to which a phase transition takes place to (no distinct thermodynamic phase below $T_g$ has been identified either experimentally or theoretically), as well as several ad-hoc assumptions of the theory \cite{dyre}. We have recently proposed that $\Delta c_p$ at $T_g$ can be understood as a purely kinetic effect \cite{heat}. In this picture, the change of $c_p$ at $T_g$ is a natural signature of the glass transition insofar as this transition is defined by the freezing of the LREs at the experimental time scale, but it is not related to a change of liquid equilibrium properties and a phase transition \cite{heat}.

In summary, we reviewed the results that follow from the recently proposed elastic approach to glass transition. This approach is based on a simple and a physically transparent idea of elastic interactions between local relaxation events in a liquid. Central to this picture is the range of this interaction, the liquid elasticity length. Its increase on lowering the temperature explains several most important open questions in the area of glass transition, including universal relaxation laws and dynamic crossovers.

There are several appealing features of the proposed theory. First, it is solely based on liquid elasticity. This is important because a glass is different from a liquid by its elastic response, namely by its ability so support shear stress. Hence, the proposed picture discusses physics that is relevant to glass transition. Second, the proposed elastic approach does not rely on ad-hoc assumptions and postulates. Often, these are not based on transparent physics and contribute to long-lasting controversies as a result \cite{dyre}. Third, the proposed picture is self-consistent. We have discussed several important phenomena, which have not been thought before to be related to one single physical mechanism, and showed that they can all be understood on the basis of elastic interactions in a liquid.


\begin{thebibliography}{99}

\bibitem{angell} C. A. Angell, K. L. Ngai, G. B. McKenna, P. F. McMillan and S. W. Martin, J. Appl. Phys. {\bf 88}, 3113 (2000).

\bibitem{dyre} J. C. Dyre, Rev. Mod. Phys. {\bf 78}, 953 (2006).

\bibitem{langer} J. Langer, Physics Today {\bf 60}, 8 (Feb. 2007).

\bibitem{ngai} K. L. Ngai, J. Non-Cryst. Sol. {\bf 353}, 709 (2007).

\bibitem{angell1} C. A. Angell, Science {\bf 267}, 1924 (1995).

\bibitem{phillips} J. C. Phillips, Phys. Rev. B {\bf 73}, 104206 (2006).

\bibitem{sti} F. Stickel, E. W. Fischer and R. Richert, J. Chem. Phys. {\bf 104}, 2043 (1996).

\bibitem{cross} J. Colmenero, A. Arbe and A. Alegria, Phys. Rev. Lett. {\bf 71}, 2603 (1993); R. Zorn et al, Phys. Rev. E {\bf 52}, 781 (1995); C. M. Roland, K. L. Ngai and L. J. Lewis, J. Chem. Phys. {\bf 103}, 4632 (1995); K. L. Ngai, C. M. Roland and G. N. Greaves, J. Non-Cryst. Sol. {\bf 182}, 172 (1995); J. Colmenero et al, Phys. Rev. Lett. {\bf 78}, 1928 (1997); R. Casalini, K. L. Ngai and C. M. Roland, Phys. Rev. B {\bf 68}, 014201 (2003).

\bibitem{casa} R. Casalini, K. L. Ngai and C. M. Roland, Phys. Rev. B {\bf 68}, 014201 (2003).

\bibitem{cross2} A. Sch\"{o}nhals, Europhys. Lett. {\bf 56}, 815 (2001); V. N. Novikov and A. P. Sokolov, Phys. Rev. E {\bf 67}, 031507 (2003).

\bibitem{e1} K. Trachenko, J. Non-Cryst. Sol. {\bf 354}, 3903 (2008).

\bibitem{e11} K. Trachenko, C. M. Roland and R. Casalini, J. Phys. Chem. B {\bf 112}, 5111 (2008).

\bibitem{e3} K. Trachenko and V. V. Brazhkin, J. Phys.: Cond. Matt. {\bf 20}, 075103 (2008).

\bibitem{e2} K. Trachenko, Phys. Rev. B. {\bf 75}, 212201 (2007).

\bibitem{elast} L. D. Landau and E. M. Lifshitz, Theory of Elasticity (Pergamon Press, 1986).

\bibitem{frenkel} J. Frenkel, Kinetic Theory of Liquids (ed. R. H. Fowler, P. Kapitza, N. F. Mott, Oxford University Press, 1947).

\bibitem{yama} O. Yamamuro et al, J. Phys. Chem. B {\bf 102}, 1605 (1998).

\bibitem{adam} G. Adam and J. H. Gibbs, J. Chem. Phys. {\bf 43}, 139 (1965).

\bibitem{nemilov} S. V. Nemilov, J. Non-Cryst. Sol. {\bf 352}, 2715 (2006).

\bibitem{dyre1} J. C. Dyre, N. B. Olsen and T. Christensen, Phys. Rev. B 53 (1996) 2171.

\bibitem{novikov} See supplementary information in V. N. Novikov and A. P. Sokolov, Nature 431 (2004) 961.

\bibitem{heat} K. Trachenko, arXiv:0807.3871v1.

\end{thebibliography}
\end{document}